\documentclass[a4paper,french]{rnti}

\usepackage[T1]{fontenc}
\usepackage[utf8]{inputenc}

\usepackage{url}

\usepackage{graphicx}

\usepackage{algorithm2e}

\usepackage[lofdepth,lotdepth]{subfig}

\usepackage{amsfonts}
\usepackage{amsmath}

\SetKwInput{KwRes}{R\'esultat}%
\SetKwIF{Si}{SinonSi}{Sinon}{si}{alors}{sinon si}{sinon}{fin si}%
\SetKwFor{Tq}{tant que}{faire}{fin tq}%
\SetKwComment{Comment}{/* }{ */}

\titrecourt{ShiftDTW : DTW pour les séries temporelles cycliques}

\nomcourt{L. Foulon et al.}

\titre{ShiftDTW : adaptation de la métrique DTW pour la clusterisation de séries temporelles cycliques}

\auteur{Lucas Foulon\affil{1},
        Ilyes Korichi\affil{1}\\
        Xavier Millot\affil{2}}

\affiliation{
    \affil{1}Data-Major\\
          prenom.nom@data-major.com,\\
          \http{https://www.data-major.com/}\\
    \affil{2}Oxtaam\\
          prenom.nom@oxtaam.com\\
          \http{https://www.oxtaam.com/}\\
 }

\resume{%
L'élasticité de la métrique DTW permet une comparaison plus souple entre les séries temporelles, et est utilisée dans de nombreux domaine de l'apprentissage automatique, comme la classification ou la clusterisation. Cependant, elle ne permet pas de faire coïncider les mesures de début et de fin de séries temporelles si elles possèdent un décalage qui intervient dès le début d'une des séries et dont la partie omise se retrouve à la fin de cette dernière. Dû à la cyclicité de ce type de séries --- qui ne possèdent ni début ni fin --- nous nous appuyons sur les travaux Cyclic DTW pour proposer une approximation moins coûteuse de cette méthode de calcul qui sera ensuite utilisée avec la méthode de clustering K-Means.
}

\summary{%
The elasticity of the DTW metric provides a more flexible comparison between time series and is used in numerous machine learning domains such as classification or clustering. However, it does not align the measurements at the beginning and end of time series if they have a shift occurring right at the start of one series, with the omitted part appearing at the end of that series. Due to the cyclicity of such series - which lack a definite beginning or end - we rely on the Cyclic DTW approach to propose a less computationally expensive approximation of this calculation method. This approximation will then be employed in conjunction with the K-Means clustering method.
}
\RestyleAlgo{ruled}

\begin{document}

%

\section{Introduction}\label{sec:intro}

Les séries temporelles sont présentes dans de nombreux contextes tel que les cours des actions boursières, les données météorologiques, ou encore les mesures biomédicales~\citep{TSreview2015Aghabozorgi}. L'enjeu du clustering est de trouver des similitudes entre ces données tout en catégorisant différemment les données les plus éloignées. L'objectif de notre projet consiste à regrouper au mieux les entreprises ayant des comportements comptables similaires. Ces comportements sont basés sur les indicateurs comptables de chaque entreprise pour trouver les meilleurs regroupements possibles. Les indicateurs comptables représentent des montants financiers, des données numériques qui évoluent dans le temps et qui permettent de mesurer la santé financière d'une entreprise. Cependant, toutes les entreprises n'ont pas la même ancienneté, et pour les comparer, nous normalisons chaque indicateur sous la forme d'une année ``type'', représentant au mieux le comportement de l'indicateur sur une année. Cette étape de normalisation peut être faite de différentes façons : dans notre cas, nous utilisons la représentation de la saisonnalité fournie par l'outil \textsc{Facebook Prophet} \citep{taylor2018forecasting} permettant ainsi d'uniformiser la durée de tous les indicateurs de toutes les entreprises. En effet, \textsc{Prophet} a été développé pour obtenir des prévisions sur les séries temporelles ayant de fortes saisonnalités (quotidienne, hebdomadaire, annuelle, etc). À partir du modèle généré, il est possible d'en extraire la tendance et les saisonnalités associées.

Mais quelle que soit la méthode d'uniformisation utilisée, nous souhaitons mettre en évidence des similitudes qui existeraient à des saisons différentes. Par exemple, une entreprise de sport d'hiver et une entreprise de sport d'été peuvent avoir des comportements comptables similaires mais décalés de 6 mois. De la même façon, certains producteurs de fruits et légumes réalisent certaines transactions selon des saisons bien précises : bien que leurs comportements similaires ne se déroulent pas sur les mêmes saisons, nous souhaiterions alors pouvoir regrouper ces producteurs. Contrairement à la méthode \textit{Cyclic Dynamic Time Warping} (CDTW) proposé par \cite{palazon2012cdtw}, nous ne souhaitons pas garder une élasticité totale : nous utilisons une bande de Sakoe-Chiba permettant de réduire les comparaisons extrêmes --- celles qui provoqueraient l'omission de nombreuses valeurs consécutives durant la comparaison entre les deux séries --- et nous n'attendons pas à connaître avec précision le meilleur alignement, mais le meilleur à quelques pas de temps près, au profit d'un allègement du temps de calcul. En effet, les méthodes comme K-Means demandent de recalculer les distances vers les nouveaux barycentres à chaque itération, et cela peut devenir assez coûteux.

Nous proposons une méthode ayant la même complexité de calcul que DTW, qui ne retournera pas l'alignement optimal, mais qui testera plusieurs alignements possibles. Ces alignement sont strictement définis par la taille de la bande Sakoe-Chiba utilisée. C'est notamment grâce à cette bande que nous limitons la complexité de notre méthode, comme présenté dans la partie \ref{sec:shift_dtw}.

\section{État de l'art}

En plus de la distance euclidienne, il existe de nombreuses mesures de distance adaptées aux séries temporelles. En effet, leur nature continue nous pousse à analyser leurs similarités de façons différentes que d’autres types de données numériques. Par exemple, nous pouvons citer la distance Kullback-Leibler \citep{liao2005clusteringTSsurvey}, la mesure de la \textit{Longest Common Subsequence} ou bien encore la méthode \textit{Dynamic Time Warping} (DTW) \citep{sakoe1971}. Cette dernière permet de donner de l’élasticité à l’axe du temps, que l’on pourrait sinon trouver trop rigide dans certains cas. Elle a été rendue populaire pour la reconnaissance de la parole avec \cite{sakoe1971, sakoe1978} puis de nouveau comme métrique de similarité pour les séries temporelles~\citep{keogh2000dtw}.

Plusieurs variantes de DTW ont été proposé depuis, pour améliorer les performances de calcul \citep{salvador2007fastdtw} ou convenir à une mesure de similarité plus adapté au contexte des séries temporelles. \cite{sakoe1978} proposent d'utiliser une bande appelée Sakoe-Chiba pour limiter la taille de la déformation de DTW, et donc de réduire le coût de calcul DTW. On retrouve aussi le \textit{Derivative Dynamic Time Warping} \citep{keogh2001ddtw} n'utilise plus la distance euclidienne pour la comparaison pair à pair, mais la différence des dérivées estimées en tout point pour chacune des séries. 
Dans notre contexte, les séries temporelles que nous manipulons n'ont ni début ni fin, car bien que temporelles, chacune de ces séries représente la saisonnalité annuelle d'un indicateur numérique. La méthode ``naïve'' consisterait à mesurer plusieurs fois les distances entre les deux séries en décalant une des deux séries pour chaque pas de temps, mais cela augmenterait considérablement le coût de calcul.
Pour la distance d'édition~\footnote{appelée aussi ``distance de Levenshtein''} (ED), \cite{maes1990ced} démontre qu'il est possible de réduire le coût en utilisant une approche ``divide-and-conquer''.
\cite{palazon2012cdtw} proposent la mesure CDTW, s'inspirant de l'algorithme de \cite{maes1990ced}, avec un coût de $\mathcal{O}(mn \times log(m))$ avec $m$ et $n$ la taille des deux séquences. Avec les solutions trouvées sur les séquences de départ, il est possible de borner la matrice des distances dans les itérations suivantes. Une fois bornée, les mêmes séquences sont comparées mais avec l'une d'elles décalée dans le temps. La méthode retournera l'alignement des deux séquences qui obtient le plus petit score DTW. Comme \cite{maes1990ced}, les auteurs prouvent formellement que leur méthode retourne le même résultat que la méthode ``naïve'', mais avec un coût moindre. Dans notre contexte d'application, nous n'avons pas besoin de cette garantie, et nous privilégions la réduction du coût de calcul à la précision du résultat.

Nous allons définir les concepts essentiels dans la partie \ref{sec:definitions} pour pouvoir présenter notre mesure de distance dans la partie \ref{sec:shift_dtw}. Enfin nous verrons les résultats obtenus dans la partie \ref{sec:expes} avant de conclure (partie \ref{sec:conclusion}).






\section{Définitions}\label{sec:definitions}

\paragraph{Série temporelle.} Une séquence ou une série temporelle $T$ de longueur $n$ est définie par une suite de valeurs $(t_0, t_1, \dots, t_{n-1})$ avec $t_i \in \mathbb{R}$.

\paragraph{Dynamic Time Warping.} Le calcul de la distance DTW entre deux séries $T$ et $S$, respectivement de longueur $m$ et $n$, se base sur la matrice des distances $M_{(T,S)}$ entre ces 2 séries tel que $M_{(T,S)}[i,j] = |t_i - s_j|$ avec $0 \leq i < m$ et $0 \leq j < n$. La complexité de calcul est $\mathcal{O}(n*m)$. L'alignement optimal est défini par le chemin de déformation le plus court depuis la matrice des distances cumulées, calculée à partir de $M_{(T,S)}$ tel que :

\begin{equation}\label{eq:def:matrice_distance_cumul}
Mcumul_{(T,S)}[i, j] =
    \begin{cases}
    M_{(T,S)}[0,0] & \text{si }i = j = 0, \\
    M_{(T,S)}[i,j] + min\begin{cases}
        Mcumul_{(T,S)}[i-1,j]\\
        Mcumul_{(T,S)}[i,j-1]\\
        Mcumul_{(T,S)}[i-1,j-1]
    \end{cases} & \text{sinon.}\\
    \end{cases}
\end{equation}

\begin{figure}
  \begin{center}
    \subfloat[Matrice des distances entre deux séries temporelles.]{
      \includegraphics[width=0.45\linewidth]{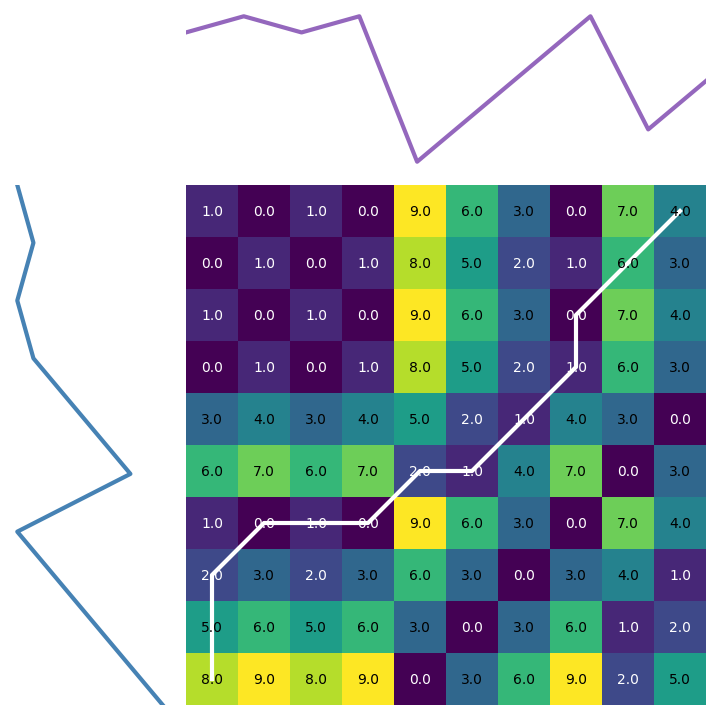}
      \label{fig:def:matrice_distance:without_mask}
                         }
    \subfloat[Matrice des distances entre deux séries temporelles avec un masque Sakoe-Chiba.]{
      \includegraphics[width=0.45\linewidth]{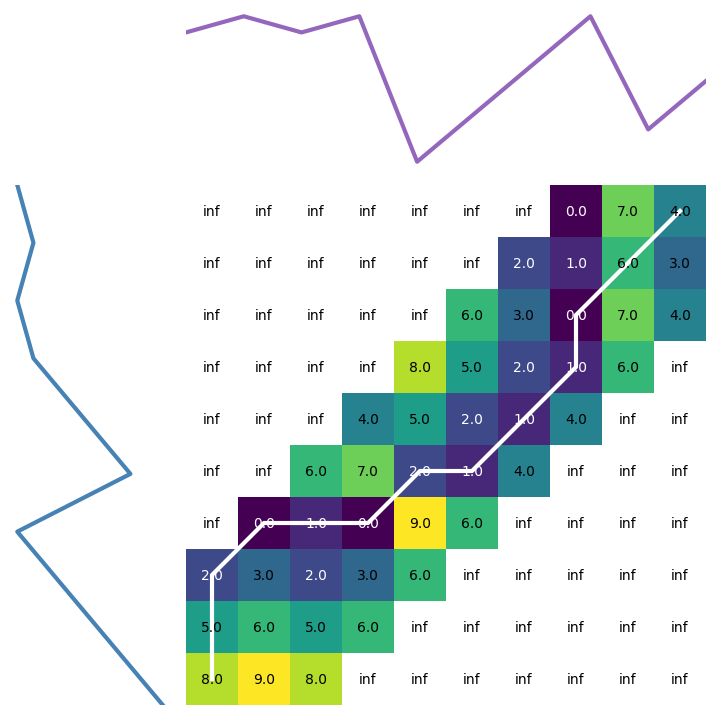}
      \label{fig:def:matrice_distance:with_mask}
                         }
    \caption{Matrices des distances entre deux séries temporelles. Les deux séries possèdent les mêmes valeurs, mais l'une d'elle a été décalé dans le temps.}
    \label{fig:def:matrice_distance}
  \end{center}
\end{figure}

La figure~\ref{fig:def:matrice_distance:without_mask} présente la matrice de distance entre deux séries temporelles, ainsi que le chemin le moins coûteux déterminé par DTW. La distance euclidienne emprunterai simplement la diagonale de cette matrice pour déterminer son résultat. Avec un masque de Sakoe-Chiba, la complexité de calcul serait réduite car nous ne parcourons plus l'ensemble de la matrice des distances, comme présenté dans la figure~\ref{fig:def:matrice_distance:with_mask}.


\paragraph{Série temporelle cyclique.} Une séquence ou une série temporelle $T$ de longueur $n$ peut subir un changement cyclique $\sigma$ tel que $\sigma(t_0, t_1, \dots, t_{n-1}) = (t_1, \dots, t_{n-1}, t_0)$.

Comme défini par \cite{palazon2012cdtw}, notons $\sigma^k$ l'arrangement d'une série temporelle de $k$ décalages, alors deux séries temporelles sont équivalentes s'il existe une valeur de $k \in \mathbb{N}$ tel que $T = \sigma^k(T')$. Alors la série temporelle cyclique $T$ est notée :
\begin{equation}
   [T] = \{\sigma^k(T) : 0 \leq k < m\}
\end{equation}






\paragraph{Cyclic Dynamic Time Warping.} La mesure CDTW \citep{palazon2012cdtw} entre $[T]$ et $[S]$ est défini tel que :

\begin{equation}
   CDTW([T], [S]) = \min_{0 \leq k < n} \biggl( \min_{0 \leq l < n} DTW(\sigma^k(T), \sigma^l(S)) \biggl)
\end{equation}

Dans leurs travaux, les auteurs démontrent qu'il est possible de réduire l'équation pour obtenir :

\begin{equation}
   CDTW([T], [S]) = \min_{0 \leq k < n} \biggl( \min (DTW(\sigma^k(T), S), DTW(\sigma^k(T)t_{k}, S))\biggl)
\end{equation}

avec $\sigma^k(T)t_{k}$ la concaténation de l'élément $t_k$ avec $\sigma^k(T)$.\\

Avec cette simplification, il propose un algorithme de type ``\textit{branch-and-bound}'' qui permet de réduire la complexité à $\mathcal{O}(mn \times log(m))$ avec $m$ et $n$ la taille des deux séquences. Ils précisent également les ce coût ne peut être réduit grâce à l'utilisation de bande de Sakoe-Chiba. Dans notre cas, nous souhaitons réduire le coût au maximum et il ne nous est pas nécessaire de tester toutes les solutions incluses dans $CDTW([T], [S])$. Nous proposons dans la partie suivante de limiter les solutions avec une bande de Sakoe-Chiba. 

\section{Présentation de la mesure ShiftDTW}\label{sec:shift_dtw}

Dans cette partie, nous allons décrire les étapes de la mesure ShiftDTW et la modification apportée à K-Means pour prendre en considération cette nouvelle mesure.

\begin{figure}
    \centering
    \includegraphics[width=0.5\linewidth]{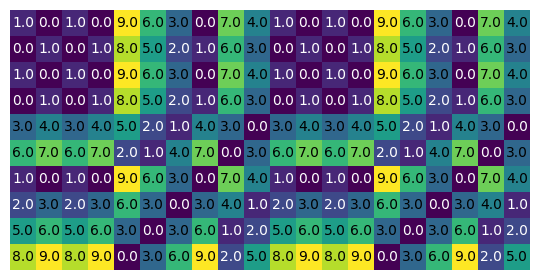}
    \caption{La matrice des distances doublée entre deux séries temporelles.}
    \label{fig:presentation:matrice_distance_x2}
\end{figure}

Notons $T$ et $S$ de longueur $m$ et $n$, deux séries temporelles à comparer. 
L'algorithme \ref{algo:shiftdtw} décrit l'ensemble des étapes nécessaires au calcul ShiftDTW entre $T$ et $S$ avec une bande Sakoe-Chiba de taille $r$.
Dans un premier temps, nous calculons les distances entre les valeurs pair-à-pair des deux séries temporelles (la matrice calculée est un des paramètres de l'algorithme \ref{algo:shiftdtw}). Nous obtenons une matrice des distances de tailles $m \times n$ que nous doublons afin d'obtenir une matrice de taille $2 \times m \times n$. Avec l'exemple de la figure~\ref{fig:def:matrice_distance:without_mask}, nous obtenons le résultat présenté dans la figure~\ref{fig:presentation:matrice_distance_x2}.

\begin{figure}
  \begin{center}
    \subfloat[Sans décalage, première itération.]{
      \includegraphics[width=0.45\linewidth]{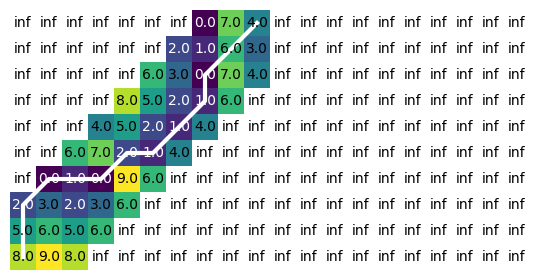}
      \label{fig:def:matrice_distance_x2_sakoechiba:ite:0}
                         }
    \subfloat[Avec un décalage, seconde itération.]{
      \includegraphics[width=0.45\linewidth]{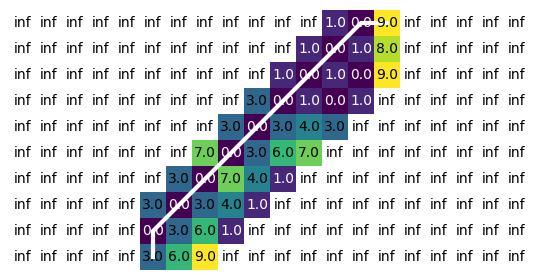}
      \label{fig:def:matrice_distance_x2_sakoechiba:ite:1}
                         }
    \caption{Matrices des distances entre deux séries temporelles avec un masque de Sakoe-Chiba. La matrice de la première itération est similaire au calcul DTW avec masque présenté dans la figure~\ref{fig:def:matrice_distance:with_mask}. La seconde itération retourne un score plus petit et donc meilleur.}
    \label{fig:def:matrice_distance_x2_sakoechiba}
  \end{center}
\end{figure}

Ensuite, comme pour DTW, la matrice des distances cumulées sera calculée comme décrit dans l'équation~\ref{eq:def:matrice_distance_cumul}, en tenant compte de la bande Sakoe-Chiba. Cette première itération est en tout point similaire à DTW classique avec la même bande Sakoe-Chiba (figure~\ref{fig:def:matrice_distance_x2_sakoechiba:ite:0}). Les itérations suivantes sont effectuées sur la même matrice de distances cumulées mais en partant avec un décalage égal à celui de la taille de la bande Sakoe-Chiba (figure~\ref{fig:def:matrice_distance_x2_sakoechiba:ite:1}). 

Pour chacune des itérations, un chemin --- le plus court --- est calculé, et la mesure ShiftDTW retourne le plus court d'entre eux avec la taille du décalage associé.

\begin{algorithm}
    \caption{Calcul de la mesure ShiftDTW entre les séries $T$ et $S$ sachant $r$.}\label{algo:shiftdtw}
    \renewcommand{\algorithmcfname}{Algorithme}%
    \SetAlgoLined
    \KwData{$M_{(T,S)}$ la matrice des distances pair-à-pair entre $T$ et $S$}
    \KwData{$r$ la taille du rayon de la bande Sakoe-Chiba}
    \KwData{$l_T$ la longueur de la série temporelle $T$}
    \KwRes{La mesure $ShiftDTW_r$ entre $T$ et $S$, et la valeur de décalage $d$}
    $ShiftDTW_r \gets \infty$\;
    $d \gets None$\;
    $M_{(2T,S)} \gets concat(M_{(T,S)}, M_{(T,S)})$ \Comment*[r]{taille $m \times n \rightarrow 2 \times m \times n$}
    $Mask \gets calcul\_masque\_Sakoe\_Chiba(r)$\;
    \Pour{$0 \leq i < l_{T}$ par pas de $2 \times r + 1$}{
        $Mcumul_{(T,S)}[:] = cumul\_depuis\_matrix\_dist(M_{(2T,S)}[i:l_T+i], Mask)$ \Comment*[r]{calcul matrice des distances cumulées}
        $current\_dist = \sqrt{Mcumul_{(T,S)}[-1, -1]}$\;
        \If{$current\_dist < ShiftDTW_r$}{
          $ShiftDTW_r \gets current\_dist$\;
          $d \gets r$\;
        }
    }
\end{algorithm}

Comme le nombre de distances cumulées visitées par l'algorithme est le même que DTW classique sans bande de Sakoe-Chiba, nous obtenons un coût de $\mathcal{O}(m \times n)$.

\paragraph{Adaptation de K-Means}

À chaque itération, l'algorithme K-Means recalcule les bary\-cen\-tres de chaque cluster. Cependant, comme l'algorithme ShiftDTW décale une des séries à chaque itération, ce décalage doit être conservé lors de la mise à jour des barycentres. Chaque série temporelle se voit attribuer un décalage qui lui est propre, selon le barycentre le plus proche et le décalage qui leur est associés.

\section{Expérimentations}\label{sec:expes}

Dans cette partie, nous allons présenter les résultats obtenus sur 2 jeux de données du domaine, puis sur un échantillon des données métiers. Dans les 2 cas, nous avons utilisé la méthode K-Means++ \citep{kmeansplusplus2007arthur} pour initialiser les barycentres.

\subsection{Jeux de données artificiels}

Nous avons d’abord testé notre méthode sur un jeu de données du domaine scientifique afin de s’assurer que notre modèle est pertinent pour répondre à notre problématique de clustering sur données cycliques. Nous avons choisi deux jeux de données, parmi la banque de données UCR time series archive \citep{UCRArchive2018}, qui sont intrinsèquement cyclique car elles sont finies et n'ayant ni début ni fin : BeetleFly et BirdChicken --- utilisés pour la première fois par \cite{hills2014classification}. Ces données représentent les formes de scarabées et de mouches pour le premier, et d’oiseaux et de poulets pour le second, et sont de longueur 512.
Nous avons testé la distance euclidienne, DTW et Shift DTW avec $r = 4$ avec l’algorithme K-Means, en initialisant 10 fois les barycentres pour ne garder que le meilleur score pour chaque mesure de distance. Ces deux jeux de données contiennent chacun 2 groupes de 10 séries temporelles. Nous présentons dans le tableau~\ref{tab:res:artificiels} les résultats de précision sur les jeux d'apprentissage (train).

\begin{table}[!h]
    \centering
    \begin{tabular}{l|ccc} \hline 
         & k-means euclidien & k-means DTW & k-means Shift DTW\\ \hline 
       BeetleFly - train  & 0.8 & 0.75 & \textbf{0.9}\\ 
       BirdChicken - train & \textbf{0.8} & 0.7 & 0.75\\ 
    \end{tabular}
    \caption{Présentation des résultats de classification sur les jeux d'apprentissage BeetleFly et BirdChicken.}
    \label{tab:res:artificiels}
\end{table}

On remarque que la mesure Shift DTW est meilleure sur BeetleFly, et est en seconde position sur BirdChicken. Notre mesure montre alors qu’elle est plus pertinente dans certains cas, et qu'elle est meilleure --- et donc tout aussi pertinente --- que la mesure DTW classique.

\subsection{Jeux de données comptables}

\begin{figure}
    \centering
    \includegraphics[width=0.7\linewidth]{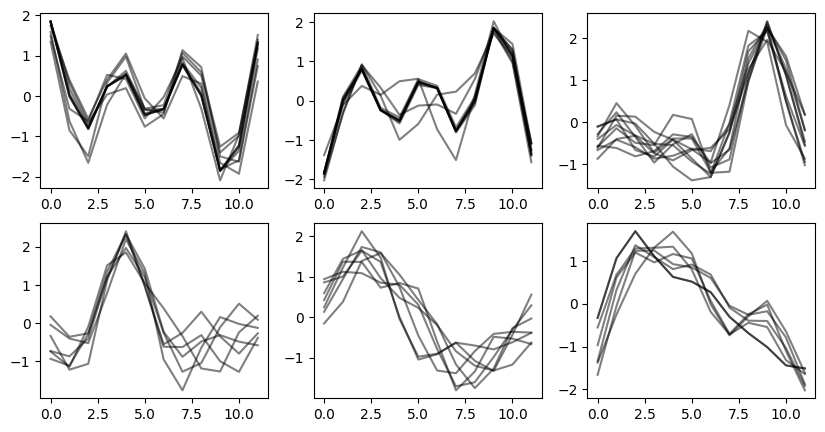}
    \caption{Présentation des 6 clusters sélectionnés parmi les 100 clusters calculés avec K-Means euclidien.}
    \label{fig:res:origin}
\end{figure}

Nous avons ensuite expérimenté notre méthode sur les données métiers. L'objectif ici est de montrer qu'il est possible de regrouper des séries temporelles qui se ressemblent mais qui possèdent un décalage dans le temps. Le décalage doit être assez grand pour que la mesure euclidienne calcule des distances trop grande pour les réunir dans un même cluster K-Mean. Parmi les 1042 séries temporelles, donc chacune d’entre elles représente une et une seule entreprise : nous avons choisi un échantillon de 50 séries temporelles pour ne pas à avoir un étiquetage manuel des séries trop lourd. Pour les choisir, nous avons d’abord appliqué un K-Means euclidien avec 100 clusters, de façon à séparer au mieux ces séries temporelles, et nous avons choisi 6 clusters présenté dans l'image~\ref{fig:res:origin} qui nous semblaient pertinents pour notre démonstration. Nous souhaitons, par exemple, que les deux premiers clusters soient réunis pour n'en former qu'un, en tenant compte de la cyclicité avec notre méthode. 

\begin{figure}[ht]
  \begin{center}
    \subfloat[Cluster 1]{
      \includegraphics[width=0.45\linewidth]{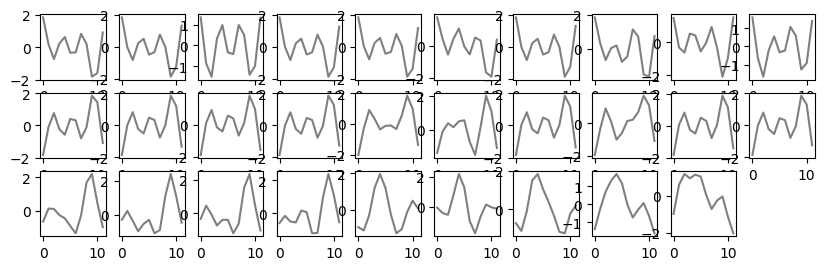}
                         }
    \subfloat[Cluster 2]{
      \includegraphics[width=0.45\linewidth]{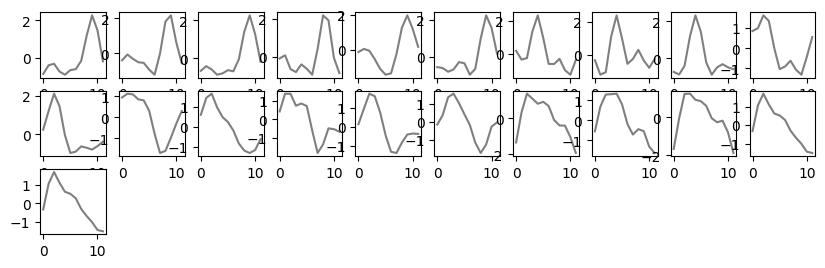}
                         }
    \caption{Résultats du clustering avec les séries temporelles de longueur 12.}
    \label{fig:res:clust12}
  \end{center}
\end{figure}

\begin{figure}[ht]
  \begin{center}
    \subfloat[Cluster 1]{
      \includegraphics[width=0.45\linewidth]{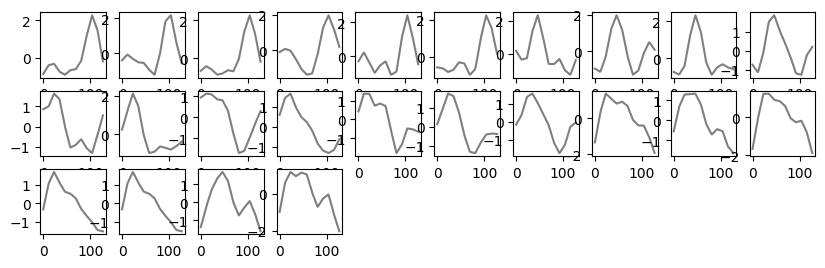}
                         }
    \subfloat[Cluster 2]{
      \includegraphics[width=0.45\linewidth]{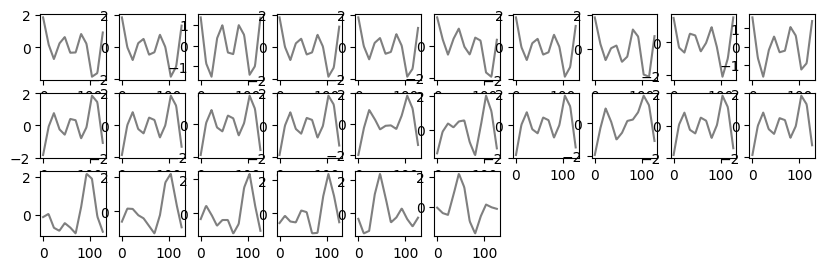}
                         }
    \caption{Résultats du clustering avec les séries temporelles de longueur 128.}
    \label{fig:res:clust128}
  \end{center}
\end{figure}

\begin{figure}[ht]
  \begin{center}
    \subfloat[Cluster 1]{
      \includegraphics[width=0.45\linewidth]{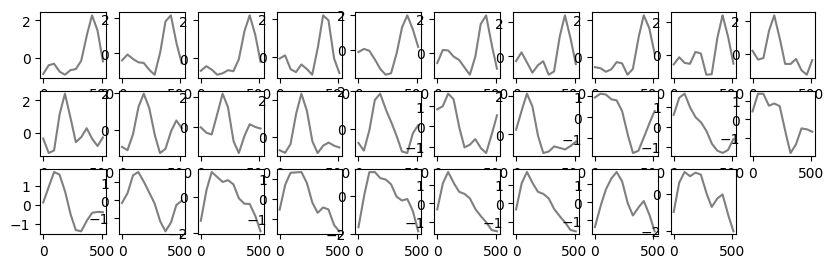}
                         }
    \subfloat[Cluster 2]{
      \includegraphics[width=0.45\linewidth]{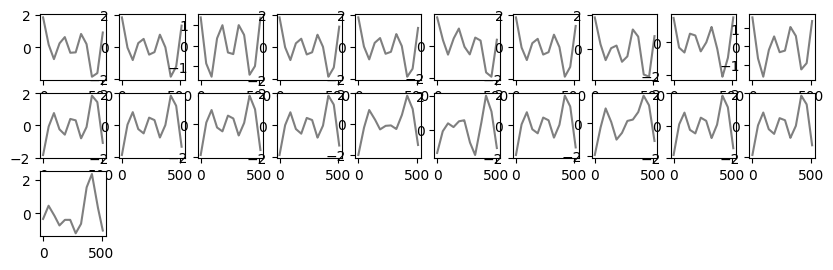}
                         }
    \caption{Résultats du clustering avec les séries temporelles de longueur 512.}
    \label{fig:res:clust512}
  \end{center}
\end{figure}

En effet, il fallait montrer que certaines séries similaires en tenant compte d’un certain décalage, sont bien regroupées avec notre méthode. Nous avons testé en encodant les séries sur différentes longueurs. Sur des séries temporelles de longueur 12 (une valeur par mois) et un paramètre de décalage à 3, les résultats sont intéressants. Par contre, en augmentant la taille des séries temporelles, et en conservant un petit décalage, les regroupements tendent davantage vers notre objectif : les séries décalées dans le temps se rassemblent mieux. C’est également pour cela que notre méthode a donné de bons résultats sur les jeux précédents. Nous pouvons observer les 2 clusters trouvés sur ces images avec des longueurs 12. Nous pouvons observer les résultats dans les images~\ref{fig:res:clust12},~\ref{fig:res:clust128} et~\ref{fig:res:clust512}.

Nous constatons que plus la taille des séries augmente, et plus le premier groupe devient petit et tend à être plus homogène dans la forme des séries qui y sont contenues. On peut toutefois supposer que le deuxième groupe pourrait peut-être se diviser en deux sous-groupes, mais nous n’avons pas tester avec 3 barycentres.
Ces résultats n’auraient pas pu être obtenus avec une distance euclidienne ou bien même avec DTW, et la méthode Cyclic DTW~\cite{palazon2012cdtw} aurait été plus coûteuse en temps de calcul pour parvenir à ces résultats.

\section{Conclusion \& Perspectives}\label{sec:conclusion}

Nous obtenons de bons résultats en terme de clustering sur ces jeux de données. Nous souhaiterions tester notre méthode sur de plus nombreux jeux de données. Nous aimerions également comparer ces résultats avec ceux de la méthode de \cite{palazon2012cdtw}.

Nous pourrions également tester d’autres algorithmes de clustering tel que Nearest Neighbors Classifier (utilisé également dans \cite{UCRArchive2018}) ou \textsc{Dbscan}~\citep{dbscan1996ester}. 
%
%
%
%
Une piste très intéressante est le calcul de barycentre adapté aux mesures DTW proposé par~\cite{bary4dtw2011petitjean}. Les auteurs montrent des gains d’inertie intra clusters et globaux très intéressants.

En ce qui concerne son usage sur les données métiers, nous souhaitons améliorer le pré-traitement en évaluant, pour chaque entreprise, le meilleur paramétrage à utiliser avec \textsc{Prophet} pour limiter l’erreur entre les chiffres d'affaire de chaque année et la saisonnalité générée.
Il est aussi envisagé d’utiliser notre méthode sur d’autres indicateurs que le chiffres d'affaire, voire de combiner les indicateurs pour travailler en multi-dimensionnel. Cela impose bien sûr d’adapter les méthodes déjà implémentées.




\bibliographystyle{rnti}
\bibliography{ref}



\Fr

\end{document}